# The thermal and electrical properties of the promising semiconductor MXene $Hf_2CO_2$

Xian-Hu ZHA[1], Qing HUANG[1], Jian HE[2], Heming HE[3], Junyi Zhai[4], Yue Wu[5], Joseph S. FRANCISCO[6], Shiyu DU[1] *

[1]Division of Functional Materials and Nanodevices, Ningbo Institute of Materials Technology and Engineering, Chinese Academy of Sciences, Ningbo, Zhejiang 315201, China

[2]Dalian Institute of Chemical Physics, Chinese Academy of Sciences, Dalian, 116023, China

[3]State Nuclear Power Research Institute, Beijing, 100029, China

[4]Beijing Institute of Nanoenergy and Nanosystems, Chinese Academy of Sciences, Beijing 10083, China

[5]Department of Chemical and Biological Engineering, Iowa State University, 2033 Sweeney Hall, Ames, IA 50011, USA

[6]Departments of Chemistry and Earth and Atmospheric Science, Purdue University, West Lafayette, IN 47906, USA

**ABSTRACT** The family of 2D transition-metal carbides and nitrides labeled as "MXenes" has recently attracted considerable attention. The diverse applications and excellent performance of MXenes have been demonstrated both experimentally and theoretically. However, very few studies have addressed the thermal and electrical properties of MXene materials. In this work, we investigate the thermal and electrical properties of oxygen-functionalized $M_2CO_2$ (M = Ti, Zr, Hf) MXenes using first-principles calculations. In this work, $Hf_2CO_2$ is found to exhibit a thermal conductivity better than $MoS_2$ and phosphorene. The room-temperature thermal conductivity along the armchair direction is determined to be 86.25~131.2 $Wm^{-1}K^{-1}$ with a flake length of 5~100 μm, and the corresponding value in the zigzag direction is approximately 42% of that in the armchair direction. Other important thermal properties of $M_2CO_2$ are also considered, including their specific heat and thermal expansion coefficients. The theoretical room temperature thermal expansion coefficient of $Hf_2CO_2$ is $6.094×10^{-6}$ $K^{-1}$, which is lower than that of most metals. Moreover, $Hf_2CO_2$ is determined to be a semiconductor with a band gap of 1.657 eV and to have high and anisotropic carrier mobility. At room temperature, the $Hf_2CO_2$ hole mobility in the armchair direction (in the zigzag direction) is determined to be as high as $13.5×10^3$ $cm^2V^{-1}s^{-1}$ ($17.6×10^3$ $cm^2V^{-1}s^{-1}$), which is comparable to that of phosphorene. Broader utilization of $Hf_2CO_2$ as a material for nanoelectronics is likely because of its moderate band gap, satisfactory thermal conductivity, low thermal expansion coefficient, and excellent carrier mobility. The corresponding thermal and electrical properties of $Ti_2CO_2$ and $Zr_2CO_2$ are also provided here for comparison. Notably, $Ti_2CO_2$ presents relatively low thermal conductivity and much higher carrier mobility than $Hf_2CO_2$, which is an indication that $Ti_2CO_2$ may be a better choice for nanodevice when heat conduction is not the major concern. This finding also makes the foundation for $Ti_2CO_2$ to be used as an efficient thermoelectric material.

Because two-dimensional (2D) materials exhibit many novel electronic and thermal properties that differ from those of their bulk counterparts, these materials have received considerable attention over the past two decades. For example, the adoption of 2D materials in place of traditional bulk materials for the next generation of electronic devices has recently been demonstrated as a potentially practical strategy[1]. Since the size of electronic devices has been continuously decreasing over the past twenty years during the development of highly integrated electronic components, efficient heat dissipation, a moderate electronic band gap and a reasonably high carrier mobility[2] have become equally important properties in determining the performance of electronic devices. Graphene, a well-known 2D carbon material[3], has been demonstrated to possess a high thermal conductivity and high charge carrier mobility[4], which have strongly encouraged ongoing research into expanding its applicability as a material for thermal conduction or electronic devices. Many other similar classes of 2D materials have also been fabricated as candidate materials for nanoelectronics, including monolayers of h-BN[5], phosphorene[2, 6-8], and transition-metal dichalcogenides[9, 10]. In-depth investigations of these monolayers have been conducted and have led to several potential applications[11]. However, additional treatment on these 2D materials is generally required, such as structural modification, composition, or application of external fields[11] for their practical application in highly integrated electronic components. The reasons for these requirements are as follows: graphene is a zero-band-gap semiconductor in nature and thus its band gap needs to be opened[12]; monolayer h-BN[13] has an excessively large band gap (5.5 eV); the thermal conductivities of monolayer $MoS_2$ and phosphorene [$(34.5\pm4)$ and 11.8 $Wm^{-1}K^{-1}$, respectively] are not impressive in terms of heat dissipation[7, 14]. Moreover, the carrier mobility of $MoS_2$ is not ideally high[15], and phosphorene is prone to chemical degradation upon exposure to ambient conditions[16]. In other words, difficulties may occur if these materials are directly applied in highly integrated electronic components considering their intrinsic properties. Therefore, the discovery of a desirable 2D semiconducting material with a moderate band gap, satisfactory intrinsic thermal conductivity and high carrier mobility remains a primary goal of research in physics and materials science.

Recently, a new family of 2D transition-metal carbides and nitrides called "MXene" [17-22] has been fabricated by selective etching of "A" from $M_{n+1}AX_n$ phases (where M is an early transition metal, A is an A-group element, X is C and/or N, and n=1, 2, or 3). Since $M_{n+1}AX_n$ materials represent a large family that consists of more than 60 members[23], the corresponding MXenes inherit versatile configurations. To date, the following MXenes have been synthesized by exfoliation of the corresponding $M_{n+1}AX_n$ phases with a hydrofluoric acid treatment: $Ti_3C_2$[17], $Ti_2C$, $Ta_4C_3$, $TiNbC$, $(V_{0.5},Cr_{0.5})_3C_2$, $Ti_3CN$[18], $Nb_4C_3$[22], $Nb_2C$ and $V_2C$[19]. The as-synthesized MXenes are typically functionalized by -O, -OH and -F groups. Naguib et al.[21] have published a review of these materials in which they denote functionalized MXenes as $M_{n+1}X_nT_x$, with T standing for the surface-terminating group. Both theoretical and experimental results have demonstrated that MXenes have potential applications in hydrogen storage[24], lithium-ion batteries (LIBs)[25, 26], supercapacitors[27, 28], adsorbents[29, 30] and electronic devices[31, 32]. For example, the 2D material $Sc_2C$ has a predicted hydrogen storage capacity of 9.0%, which is much higher than the target hydrogen gravimetric storage capacity (5.5 wt.% by 2015) set by the U.S. Department of Energy (DOE)[24]. $Ti_3C_2$ has been demonstrated to be a promising anode material for LIBs from a first-principles perspective[26]. The results from experiments on 2D $Nb_2C$- and $V_2C$-based electrodes indicate reversible capacities of 170 and 260 mAhg$^{-1}$ at 1C, respectively, and 110 and 125 mAhg$^{-1}$ at 10C, respectively[19]. MXenes also have a high capacity for cations[27] (e.g., $Na^+$, $K^+$, $NH_4^+$, $Mg^{2+}$, and $Al^{3+}$), pollutant ions[29] ($Pb^{2+}$), and small molecules[30] ($N_2H_2$), with applications that include supercapacitors and water-pollution abatement processes. Moreover, MXenes have been recently applied to the design of electronic nanodevices combined with transition-metal dichalcogenide monolayers[31, 32]. As the attention on this new class of MXene materials grows, gaining deeper insight into their basic physical properties becomes more important. However, very few studies have been published in the literature on the thermal properties and carrier mobility of MXene materials.

In this work, the electronic band gap, thermal properties and carrier mobility of three oxygen-functionalized MXenes, $Ti_2CO_2$, $Zr_2CO_2$ and $Hf_2CO_2$, are predicted via theoretical calculations. Oxygen-functionalized MXenes are chosen in this work for their current applicability compared to those functionalized by -F and -OH groups[26, 33-36] and for their higher thermodynamic stability[33]. Based on the results presented herein, $Hf_2CO_2$ is unexpectedly determined to possess a moderate band gap, a high carrier mobility comparable to phosphorene, and a thermal conductivity better than $MoS_2$ and phosphorene, which indicates that this material may have extensive potential applications in nanoelectronics. For this reason, the major part of this work focuses on the electrical properties and thermal conductivity of $Hf_2CO_2$. The corresponding values for $Ti_2CO_2$ and $Zr_2CO_2$ are also examined and discussed for comparison to the results for $Hf_2CO_2$. According to the computational results, the thermal conductivity increases with increasing atomic number of M among $M_2CO_2$ (M=Ti, Zr, Hf) MXenes and all the three MXenes present high carrier mobilities. Moreover, $Ti_2CO_2$ can be an efficient thermoelectric material with much lower thermal conductivity and higher hole mobility than $Zr_2CO_2$ and $Hf_2CO_2$. The results for $Ti_2CO_2$ and $Zr_2CO_2$ are mainly supplied in the Supporting Information. In addition, the specific heat and thermal expansion coefficient of $Hf_2CO_2$ are provided as well.

**Computational Details.** The calculated structures and electronic properties are determined on the basis of first-principles density functional theory implemented in the plane-wave VASP code[37]. The generalized gradient approximation (GGA) of the Perdue-Burke-Ernzerhof (PBE)[38] scheme is adopted for the exchange-correlation functional. To obtain a more reliable band gap, the Heyd-Scuseria-Ernzerhof (HSE06)[39, 40] hybrid functional is utilized to calculate the electronic energy bands of $M_2CO_2$ (M=Ti, Zr, Hf) hexagonal unit cells. The projected augmented wave (PAW) approach[41] is employed for pseudopotentials; the plane-wave cutoff energy is chosen to be 500 eV. The conjugate gradient[42] method is applied for structural optimization, and the system is relaxed until the forces on each atom are less than $1.0 \times 10^{-4}$ eV/atom. To eliminate neighboring layer interaction, a 25-Å vacuum layer parallel to the surface layer is used. During optimization, a $12 \times 12 \times 1$ k-points mesh is sampled in the Brillouin zone (BZ), and a 60 k-points grid is applied for plotting the electronic energy band. All of the structures are visualized using the VESTA code[43].

The thermal properties, including the phonon thermal conductivity, specific heat, and thermal expansion coefficient, are calculated from the phonon dispersion of a hexagonal unit cell, as circled in the pink rhombus in Figure 1A. From Figure 1B, which represents the hexagonal BZ, it can be found that the ΓM (ΓK) direction in the reciprocal space corresponds to the armchair (zigzag) direction of $M_2CO_2$ (M=Ti, Zr, Hf) in the real space. A 120 k-points grid is employed for plotting the phonon dispersion for various directions and the entire BZ. The Phonopy software[44] combined with the VASP code is utilized for phonon dispersion calculations. The theoretical calculation is performed with the density functional perturbation theory (DFPT)[45], and a $6 \times 6 \times 1$ k-points mesh based on a $4 \times 4 \times 1$ supercell is adopted for calculating the dynamical matrix. The phonon band connections are estimated from eigenvectors and the phonon band is determined considering band crossings implemented in Phonopy software[44]. The phonon thermal conductivity is calculated as follows[46]:

$$\kappa_{ph} = \sum_j \int C_j[\omega_j(q)] v_j(q) l_j(q) dq \qquad (1),$$

where q is the wave vector in the irreducible BZ. The contribution to the heat capacity of the $j^{th}$ phonon branch is denoted by $C_j[\omega_j(q)]$, and $\omega_j(q)$ is the circular phonon frequency. The term $v_j(q)$ represents the group velocity along the temperature gradient, which is calculated by $v_j(q) = d\omega_j(q)/dq$. The term $l_j(q)$ is the phonon mean free path. Within the framework of Klemens' theory[47, 48], Equation (1) can be redefined as:

$$\kappa_{ph} = \sum_j \frac{\rho}{<\gamma_j^2>} \frac{<v_j>^4}{\omega_{max,j} T} \ln \frac{\omega_{max,j}}{\omega_{min,j}} \qquad (2),$$

where T is temperature, and $k_B$ and $\hbar$ are the Boltzmann and reduced Plank's constants, respectively. $\omega_{max,j}$ and $\omega_{min,j}$ are the maximum and minimum circular frequency of each $j^{th}$ branch. Due to the finite flake length $L$, the term of $\omega_{min,j}$ is redefined as[48]:

$\omega_{min,j} = (\frac{M<v_j>^3 \omega_{max,j}}{2<\gamma_j^2> k_B T L})^{\frac{1}{2}}$, with $M$ being the mass of the MXene unit cell. $\gamma_j$ is the average value of the $j^{th}$ branch Grüneisen parameter, which is given by[49] $\gamma_j(q) = -\frac{a}{2\omega_j(q)} \frac{\partial \omega_j(q)}{\partial a}$.

$<\gamma_j^2>$ in Equation (2) is estimated by[47] $<\gamma_j^2> = \frac{\sum_k \gamma_{j,k}^2 C_{j,k}}{\sum_k C_{j,k}}$. Variable $\rho$ represents the mass density. For our hexagonal lattices, mass density is calculated as $\rho = M / (\frac{\sqrt{3}}{2} a^2 h)$, where $a$ is the lattice parameter in the *xy* plane, and $h$ is the layer thickness. The value of *h* is calculated by the distance between two neighboring carbon bilayers of $M_2CO_2$ MXene (6.883 Å for $Ti_2CO_2$; 6.191 Å for $Zr_2CO_2$; 6.029 Å for $Hf_2CO_2$), which is similar to the calculation of graphene[50]. To accurately describe the interlayer interaction of the bilayers, a damped van der Waals (VDW) correction (DFT-D2)[51] is adopted. The flake length, $L$, which ranges from 1 to 100 μm, approaching the experimental results[28] are considered. The thermal conductivities along the armchair and zigzag directions are both investigated, similar to the previous study conducted for graphene monoxide[52]. The method for thermal conductivity is verified by calculating two remarkably different thermal conductivities of graphene and ZnSb (4755.6 and 4.227 Wm$^{-1}$K$^{-1}$ for graphene and ZnSb respectively, based on a 5 μm flake length at room temperature). The predicted values are consistent very well with those experimental results[4, 53]. The specific heat[54] and thermal expansion coefficient[49, 55] are calculated according to existing methods and our previous work.

The carrier mobilities of $M_2CO_2$ (M=Ti, Zr, Hf) are calculated using an orthorhombic unit cell, as circled in the blue box in Figure 1A. The $x$-direction in the real-space corresponds to the zigzag direction of $M_2CO_2$ MXenes, and the $y$-direction is along the armchair direction. The corresponding BZ of the orthorhombic cell is presented in Figure 1C. The carrier mobility is calculated according to Equation (3), considering electron-phonon coupling[2, 8, 56].

$$\mu = \frac{e\hbar^3 C}{k_B T m_e^* m_d (E_1^i)^2} \tag{3}$$

here $m_e^*$ is the carrier effective mass along the transport direction; $m_d$ is determined by $m_d = \sqrt{m_{ex}^* m_{ey}^*}$, where $m_{ex}^*$ and $m_{ey}^*$ are the effective mass along the $x$- and $y$-directions, respectively; $E_1^i$ is the deformation potential constant of the valance-band maximum for holes or conduction-band minimum for electrons along the transport direction, calculated by $E_1^i = \Delta V_i / (\Delta a / a_0)$ with $\Delta V_i$ as the energy change (relative to the vacuum level) of the $i^{th}$ energy band under a small lattice variation $\Delta a$ and $a_0$ as the lattice constant along the transport direction; $C$ is the elastic modulus along the transport direction, determined by extrapolation based on the relationship of $C(\Delta a/a)^2/2 = (E-E_0)/S_0$, where $(E-E_0)$ is the change of the total energy under a varying lattice constant with a small step size ($\Delta a/a \sim 5\%$) and $S_0$ is the area of the lattice in the $xy$ plane.

**Results and Discussion.** The geometric and electronic properties of $Hf_2CO_2$ are investigated using DFT calculations. The top-view and side-view of the $Hf_2CO_2$ structures are shown in Figures 1A and 1D, respectively. As seen in the side-view, a central carbon monolayer is sandwiched between two Hf layers, and the oxygen layer is directly projected to the bottom Hf layer on both sides. The stable structure is similar to those of other stable oxygen-functionalized $M_2CO_2$ (M is a transition metal) MXenes[57], including $Ti_2CO_2$ and $Zr_2CO_2$, whose geometries differ only slightly from $Hf_2CO_2$ in bond lengths and lattice constants. Figure 1E depicts the electronic energy band of $Hf_2CO_2$. From GGA calculations, $Hf_2CO_2$ is an indirect semiconductor with a band gap of 1.021 eV, which agrees well with previous findings[57]. After HSE06 correction, this band gap increases to 1.657 eV, which is comparable to those of monolayer $MoS_2$ and phosphorene[2, 9]. Since the HSE method is demonstrated to yield a -0.3 eV mean absolute error for semiconductors' band gaps[58], the true band gap for $Hf_2CO_2$ is expected in the range of 1.657~1.957 eV. All of the $M_2CO_2$ electronic band structures based on GGA calculations are provided in Figure S1 (Supporting Information). Only five $M_2CO_2$ (M = Sc, Ti, Zr, Hf, W) MXenes are semiconductors; their electronic energy bands are corrected by HSE06 functional. Due to its semiconducting behavior, the electronic thermal conductivity in $Hf_2CO_2$ is negligible.

The $Hf_2CO_2$ phonon dispersions along the armchair ($\Gamma M$) and zigzag ($\Gamma K$) directions are shown in Figures 2A and 2B, respectively. The out-of-plane acoustic (ZA), longitudinal acoustic (LA) and transversal acoustic (TA) modes are denoted with black squares, red circles and blue triangles, respectively. The ZA mode nearly coincides with the TA mode in the armchair direction, differing from that along the zigzag direction. From the phonon dispersion, the group velocities of the acoustic modes in the armchair direction are determined as $v_{ZA}=1.826\times10^3$ $ms^{-1}$, $v_{TA}=1.919\times10^3$ $ms^{-1}$ and $v_{LA}=2.065\times10^3$ $ms^{-1}$, and the corresponding group velocities in the zigzag direction are $v_{ZA}=1.641\times10^3$ $ms^{-1}$, $v_{TA}=2.075\times10^3$ $ms^{-1}$ and $v_{LA}=1.656\times10^3$ $ms^{-1}$. The value of $\gamma_j$ representing MXene anharmonic effect is determined from the phonon dispersions using optimized (a = 1.00 $a_0$ where $a_0$ is the optimized hexagonal lattice parameter in the plane parallel to BZ) and strained ($a=0.99a_0$ and $a=1.01a_0$) configurations. By definition[49], $\gamma_j$ for the acoustic modes of $Hf_2CO_2$ are $\gamma_{ZA}=-0.164$, $\gamma_{TA}=-1.254$ and $\gamma_{LA}=1.032$ in the armchair direction and $\gamma_{ZA}=-0.263$, $\gamma_{TA}=-0.916$ and $\gamma_{LA}=1.240$ in the zigzag direction. The values of $<\gamma_j^2>$ as well as $\gamma_j$ and $\upsilon_j$ used for calculating the thermal conductivities of $M_2CO_2$ (M=Ti, Zr, Hf) MXenes are given in Table 1. Generally, $\upsilon_j$ and $\gamma_j$ decrease with the increasing atomic number of M among $M_2CO_2$ (M=Ti, Zr, Hf) MXenes. The decrease of $\gamma_j$ can be attributed to the fact that the $M_2CO_2$ (M=Ti, Zr, Hf) MXene with higher atomic number of M possesses higher mechanical strength[36] (see the discussion on the relationship between $\gamma_j$ and mechanical strength in the Supporting Information). The $Hf_2CO_2$ thermal conductivities along the armchair and zigzag directions are calculated using Equation (2); the results are depicted in Figures 2C and 2D, respectively. A flake length of 5 μm is adopted in both cases. According to the figures, the $Hf_2CO_2$ thermal conductivity is strongly anisotropic. In the armchair direction, the room temperature thermal conductivity is determined to be 86.25 $Wm^{-1}K^{-1}$, the corresponding value in the zigzag direction is only 42.3% of the former. In both directions, the thermal conductivity is mainly contributed by ZA and LA modes, with the contribution from ZA mode slightly higher. For example, in the armchair direction, the ZA and LA modes' contributions to the thermal conductivity at room temperature are 48.55 and 37.58 $Wm^{-1}K^{-1}$, respectively. However, the large contribution of ZA mode to thermal conductivity is absent in $Ti_2CO_2$ and $Zr_2CO_2$, as shown in Figure S2 and Figure S3, respectively. For these two $M_2CO_2$ (M=Ti, Zr) MXenes, their conductivities are mainly contributed by TA and LA modes.

For comparison, we plot the thermal conductivities of all the three $M_2CO_2$ (M=Ti, Zr, Hf) MXenes in Figure 3 based on a 5μm flake length. Figure 3A shows the thermal conductivities in the armchair direction, and Figure 3B depicts those in the zigzag direction. Evidently, all the three MXenes present strong anisotropic thermal conductivities, and the anisotropy is more marked with a heavier M atom. For instance, at room temperature, the ratios between the zigzag and armchair thermal conductivities of $M_2CO_2$ (M=Ti, Zr, Hf) MXenes are 54.4%, 46.2% and 42.3% respectively. One can also find that the thermal conductivity increases significantly with increasing atomic number of M in both directions. As an example, the room temperature thermal conductivities of $Ti_2CO_2$ in armchair direction is determined to be 21.88 $Wm^{-1}K^{-1}$, while the corresponding values in $Zr_2CO_2$ and $Hf_2CO_2$ increase to 61.93 and 86.25 $Wm^{-1}K^{-1}$, respectively. As to the zigzag direction, the room temperature values for $M_2CO_2$ (M=Ti, Zr, Hf) MXenes follows the similar trend, which are 11.91, 28.59, 36.51 $Wm^{-1}K^{-1}$ respectively. The increasing thermal conductivity of $M_2CO_2$ (M=Ti, Zr, Hf) MXenes with a heavier M atom can be explained by the weaker anharmonic effect due to higher mechanical strength[36]. The thermal conductivity of $Hf_2CO_2$ in armchair direction is higher than many well-known thermal conductive materials, such as pure iron[59], suggesting that $Hf_2CO_2$ may have satisfactory performance in heat conduction as a 2D oxide material. Additionally, it is worthy accentuating that the thermal conductivity of $Hf_2CO_2$ is higher than $MoS_2$ and phosphorene which are well known promising semiconducting 2D materials for nanoelectronics.

Because of boundary scattering, the thermal conductivity is dependent upon the flake length $d$. The theoretical thermal conductivity of $Hf_2CO_2$ with flake lengths from 1 to 100 μm in the armchair and zigzag directions are shown in Figures 4A and 4B, respectively. The thermal conductivity increases monotonically with increasing flake length in both directions. Moreover, the thermal conductivity is more sensitive to the flake length at low temperatures. At room temperature, the thermal conductivity in the zigzag direction increases from 27.63 to 53.03 $Wm^{-1}K^{-1}$, for a flake-length increase from 1 to 100 μm. With the same range of the flake length, the room temperature thermal conductivity in the armchair direction ranges from 62.12 to 131.2 $Wm^{-1}K^{-1}$. This value further confirms the capability of $Hf_2CO_2$ for heat dissipation if used in an electronic device.

The specific heat and thermal expansion coefficient are also studied in the hexagonal BZ of $Hf_2CO_2$ (see Figure 5A). The specific heat and thermal expansion coefficient are presented in Figures 5B and 5C, respectively. The results show that the specific heat and thermal expansion coefficient increase with increasing temperature and that the room-temperature values are $0.238\times10^3$ $Jkg^{-1}K^{-1}$ and $6.094\times10^{-6}$ $K^{-1}$, respectively. The room-temperature thermal expansion coefficient is lower than that for most metals, such as $16.50\times10^{-6}$ $K^{-1}$ for bulk copper. The low thermal expansion coefficient is another advantage of $Hf_2CO_2$ in applications requiring structural stability at varying temperatures such as in electronic devices.

In the calculation of carrier mobility, the orthorhombic cell is adopted, which makes the elastic modulus along the transport direction well defined. Moreover, using this approach, the conduction-band minimum is folded to the $\Gamma$ point, as shown in Figure 1F, which facilitates the determination of the electron effective masses in the *x*- and *y*-directions. Before predicting the carrier mobility of $Hf_2CO_2$, we perform a benchmark calculation on the electron mobility of phosphorene. The room-temperature electron mobility of phosphorene is determined to be $1.387 \times 10^3$ and $0.177 \times 10^3$ cm$^2$V$^{-1}$s$^{-1}$ along the armchair and zigzag directions, respectively, which is in good agreement with the results of previous works[2, 8]. In this study, both the electron and hole mobilities of $Hf_2CO_2$ are evaluated and the results are listed in Table 2. The electron mobilities along the zigzag (*x*-) and armchair (*y*-) directions are $0.329 \times 10^3$ and $0.077 \times 10^3$ cm$^2$V$^{-1}$s$^{-1}$, respectively, at room temperature. Because two sub-bands overlap at the valance maximum, as shown in Figure 1F, with an energy difference of only 0.6 meV at the $\Gamma$ point (the corresponding difference at the $\Gamma$ point is 1.8 meV in the case of $Ti_2CO_2$, and 0.3 meV for $Zr_2CO_2$), the hole mobilities of both sub-bands should be calculated. The total hole mobility can be estimated as the statistical average of hole mobilities from the two sub-bands on the basis of Boltzmann's distribution. For clear elaboration, we denote the sub-band with a higher (lower) energy at the $\Gamma$ point as the "upper" ("lower") band in Table 2. From the "upper" sub-band (denoted in blue in Figure 1F), the hole mobilities are determined to be $0.924 \times 10^3$ and $26.0 \times 10^3$ cm$^2$V$^{-1}$s$^{-1}$ along the $Hf_2CO_2$ zigzag and armchair directions, respectively. Correspondingly, the hole mobilities of the "lower" band (denoted in megaton in Figure 1F) are calculated as $34.3 \times 10^3$ and $1.00 \times 10^3$ cm$^2$V$^{-1}$s$^{-1}$ along the zigzag and armchair directions of $Hf_2CO_2$, respectively. The high hole mobilities are mainly caused by the small hole effective mass and low deformation potential constant, as described in the table. Notably, the effective masses and deformation potential constants for the two sub-bands are close to each other in value, but their directions appear to be opposite. For the "upper" sub-band, the hole effective mass and deformation potential constant in the zigzag direction are both approximately threefold higher than those in the armchair direction, causing the highly anisotropic hole mobility (approximately twenty-eight-fold higher in the armchair direction). However, the sequence of hole mobilities along the two directions for the "lower" sub-band are totally reversed reflected by the seemingly "exchanged" effective masses and deformation potential constants. Consequently, the average hole mobilities exhibit only slight anisotropy, which are $17.6\times10^3$ and $13.5\times10^3$ $cm^2V^{-1}s^{-1}$ along the $Hf_2CO_2$ zigzag and armchair directions, respectively. The predicted high carrier mobilities are much higher than that of monolayer $MoS_2$,[15] and they comparable to the hole mobility of monolayer phosphorene[2] along the zigzag direction. Moreover, the oxygen functionalized $Hf_2CO_2$ MXene may show higher stability than phosphorene. Thus, the high carrier mobility indicates that $Hf_2CO_2$ may perform well as a material in nanoelectronics.

The carrier mobilities of $Ti_2CO_2$ and $Zr_2CO_2$ are also calculated, as shown in Table S1 and Table S2, respectively. It is noteworthy that $Ti_2CO_2$ presents much higher hole mobility compared with $Zr_2CO_2$ and $Hf_2CO_2$. Again, two sub-bands with close energy maxima (differing by 1.8 meV) at the $\Gamma$-point appear; thus, hole mobility calculations for both sub-bands are performed. Strikingly, the statistical average hole mobility of $Ti_2CO_2$ is predicted to be $33.6\times10^3$ and $26.6\times10^3$ $cm^2V^{-1}s^{-1}$ along the zigzag and armchair directions, respectively, which are approximately two fold greater than that of $Hf_2CO_2$. This observation implies that $Ti_2CO_2$ may be preferred in thermoelectric materials because of the low thermal conductivity and high carrier mobility determined here. In addition, Khazaei *et al.* have demonstrated $Ti_2CO_2$ possesses a large Seebeck coefficient[60]. Furthermore, $Ti_2CO_2$ may be preferred over $Hf_2CO_2$ for use in electronic devices in cases heat dissipation is not the major concern.

The combination of the results for the carrier mobilities with those for the thermal properties suggests that $Hf_2CO_2$ is a good choice for nanoelectronics applications. Because of the limited data currently available, further research should be conducted on the synthesis of $M_2CO_2$ (M=Ti, Zr, Hf) monolayers, and their intrinsic thermal and electrical properties should be experimentally measured. Considering the successful fabrication of $Ti_2CT_2$ (T=-F, -OH) and the existence of MAX phases $M_2AC$ (M= Zr, Hf; A=In, Tl, Sn, Pb, S)[61], there is a great anticipation on the synthesis of $M_2CO_2$ (M=Ti, Zr, Hf) using the reported preparation methods such as etching of their parental MAX phase and heat treatment on hydroxyl functionalized MXenes. We look forward to more findings from these three MXenes in the forthcoming experimental studies.

**Conclusions.** In the present work, the thermal and electrical properties of $Hf_2CO_2$ are investigated. The $Hf_2CO_2$ band gap is determined to be 1.657 eV. The thermal conductivity of $Hf_2CO_2$ in the armchair direction at room temperature is predicted to be 86.25 $Wm^{-1}K^{-1}$ with a flake length of 5 μm; this thermal conductivity is higher than those of pure iron and some other well known two dimensional materials including $MoS_2$ and phosphorene. Moreover, the $Hf_2CO_2$ thermal conductivity is anisotropic with the thermal conductivity in the zigzag direction only 42.3% of that in the armchair direction at room temperature. In addition, the thermal expansion coefficient of $Hf_2CO_2$ is lower than that of most metals. The carrier mobility of $Hf_2CO_2$ is also predicted, with consideration of electron-phonon coupling. The room-temperature hole mobility in the armchair (zigzag) direction is calculated to be as high as $13.5 \times 10^3 cm^2V^{-1}s^{-1}$ ($17.6 \times 10^3 cm^2V^{-1}s^{-1}$). Therefore, $Hf_2CO_2$ can be considered as candidate 2D materials for the design of next-generation electronic devices. The carrier mobility of $Ti_2CO_2$ is determined to be two fold higher than that of $Hf_2CO_2$ while the thermal conductivity is much lower. According to the current results, $Ti_2CO_2$ can be considered as candidate 2D thermoelectric materials and it may also be a better option than $Hf_2CO_2$ for nanoelectronics if good heat dissipation can be achieved in a device. Finally, options for further explorations of MXenes are raised on the basis of the results from the present work.

**Figures and Tables**

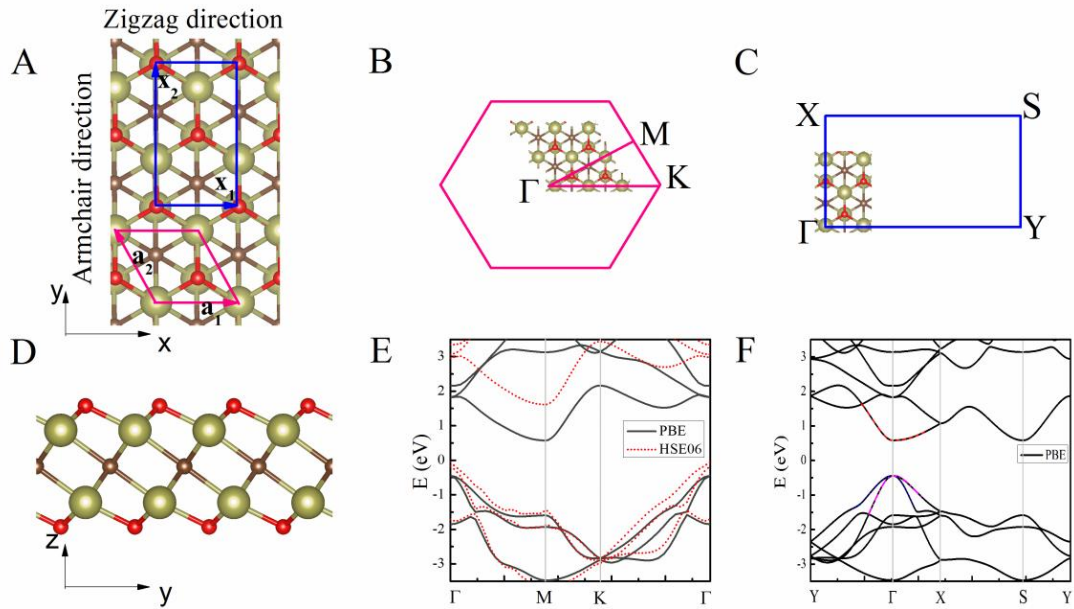

**Figure 1.** Structure and electronic band structure of $Hf_2CO_2$. The structures of $Ti_2CO_2$ and $Zr_2CO_2$ are very similar to that of $Hf_2CO_2$, except their structure have different bond lengths and lattice constants. (A) Top-view of the $Hf_2CO_2$ structure; the hexagonal unit cell and orthorhombic cell are circled in pink and blue boxes, respectively; the *x*- (*y*-) axis corresponds to the $Hf_2CO_2$ zigzag (armchair) direction. (B) The Brillouin zone of the hexagonal unit cell; the $\Gamma M$ ($\Gamma K$) direction in reciprocal space corresponds to the $Hf_2CO_2$ armchair (zigzag) direction in real space. (C) The Brillouin zone of the $Hf_2CO_2$ orthorhombic cell. (D) The side-view of $Hf_2CO_2$. (E) The electronic band structure of $Hf_2CO_2$. The band gap is increased using the HSE06 correction. The Fermi level is located at 0 eV. (F) The $Hf_2CO_2$ electronic structure based on the orthorhombic cell. The valance maximum and conduction minimum are denoted by colored lines.

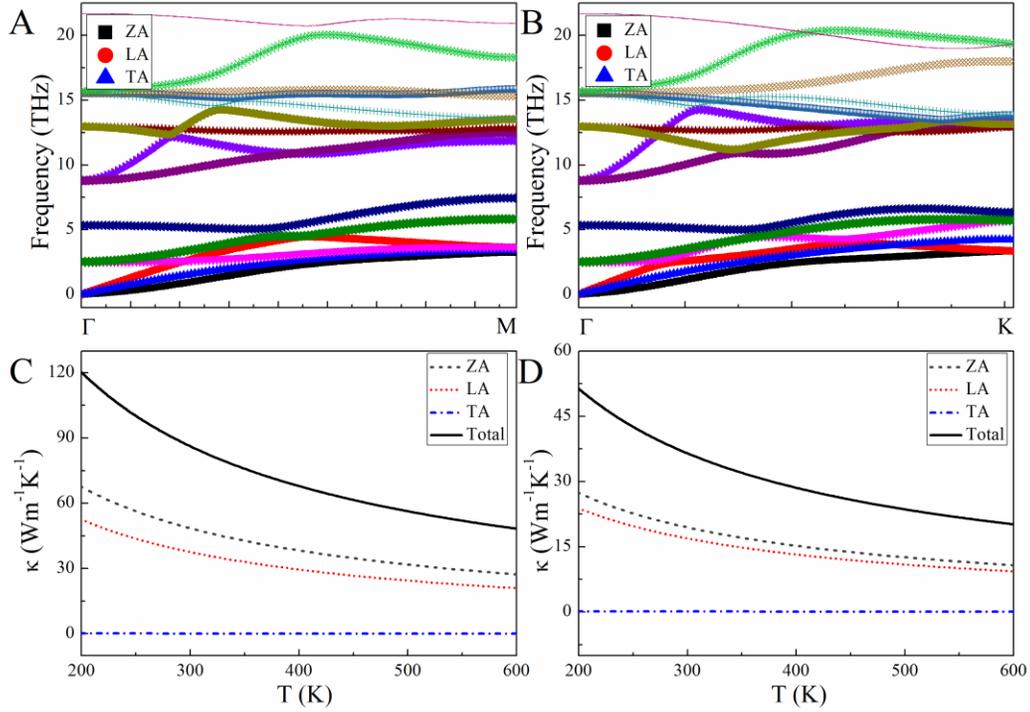

**Figure 2.** Phonon dispersions and thermal conductivities along the armchair (ΓM) and zigzag (ΓK) directions. (A) The phonon dispersion of $Hf_2CO_2$ along the armchair direction. The out-of-plane acoustic (ZA), longitudinal acoustic (LA) and transversal acoustic (TA) modes are denoted with black squares, red circles and blue triangles, respectively. (B) The phonon dispersion of $Hf_2CO_2$ along the zigzag direction. (C) The temperature dependence of the $Hf_2CO_2$ thermal conductivity along the armchair direction. The ZA, LA and LA mode contributions to the thermal conductivity are denoted with grey dashed, red dotted and blue dash-dotted lines, respectively. (D) The temperature dependence of the $Hf_2CO_2$ thermal conductivity along the zigzag direction.

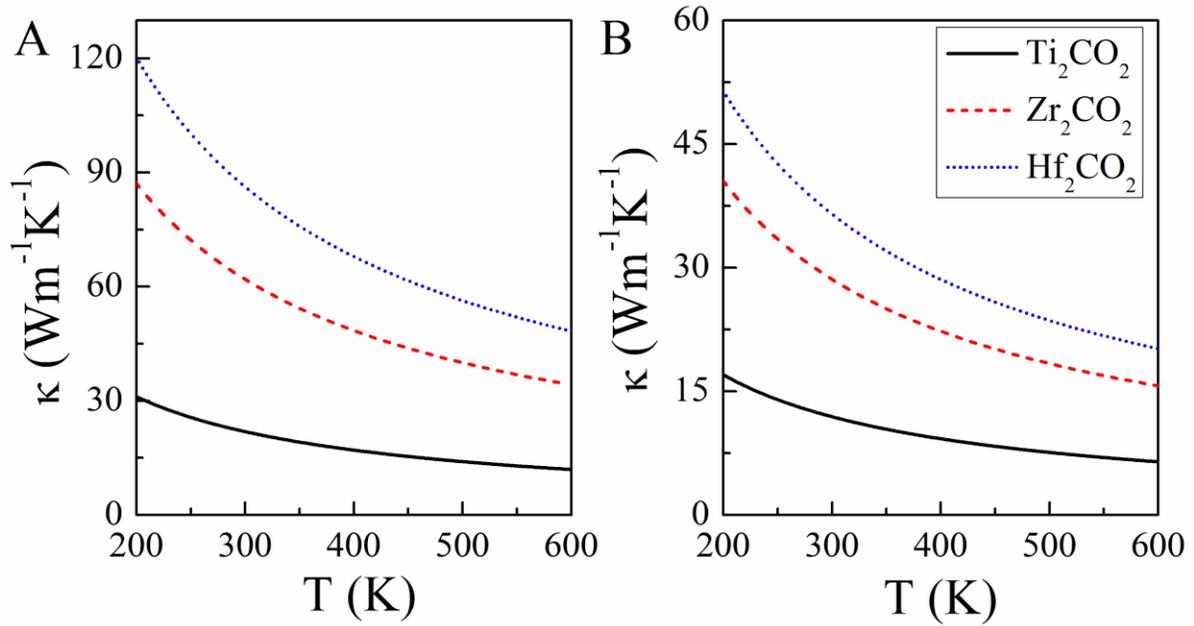

**Figure 3.** The temperature dependence of the thermal conductivities of the $M_2CO_2$ (M=Ti, Zr, Hf) MXenes. (A) The temperature dependence of the thermal conductivities of the $M_2CO_2$ (M=Ti, Zr, Hf) MXenes along the armchair direction. The $Ti_2CO_2$, $Zr_2CO_2$ and $Hf_2CO_2$ thermal conductivities are denoted in black solid, red dashed and blue dotted lines, respectively. (B) The temperature dependence of the thermal conductivities of the $M_2CO_2$ (M=Ti, Zr, Hf) MXenes along the zigzag direction.

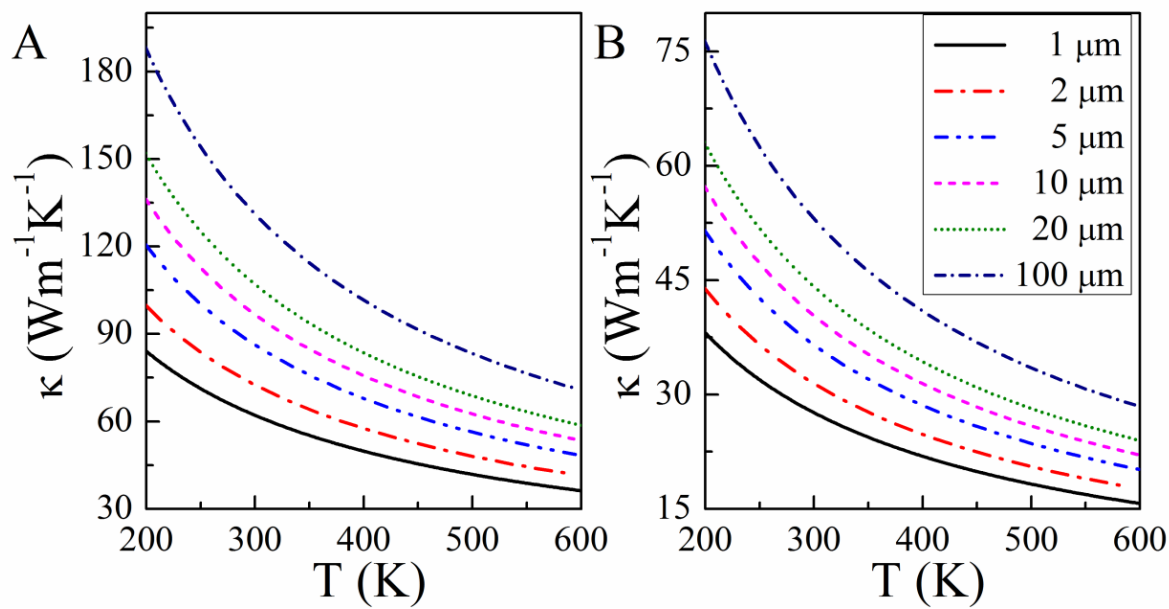

**Figure 4.** The temperature dependence of $Hf_2CO_2$ thermal conductivity with varying flake lengths. (A) The temperature dependence of the thermal conductivity with varying flake lengths in the armchair direction. The thermal conductivity for flake lengths of 1, 2, 5, 10, 20 and 100 μm are denoted by black solid, red dash-dotted, blue dashed-dotted, magenta dashed, olive dotted and navy dash-dotted lines, respectively. (B) The temperature dependence of the $Hf_2CO_2$ thermal conductivity with varying flake lengths in the zigzag direction.

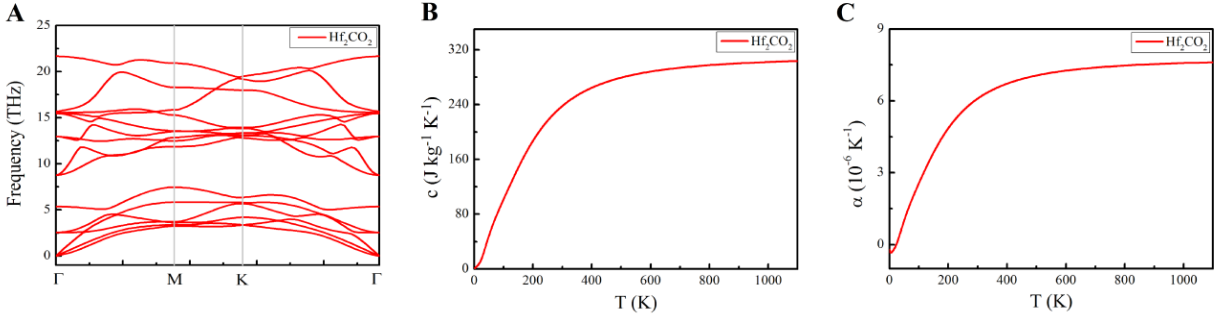

**Figure 5.** The phonon dispersion, specific heat and thermal expansion coefficient of $Hf_2CO_2$. (A) The phonon dispersion of $Hf_2CO_2$ in the Brillouin zone. (B) The temperature dependence of $Hf_2CO_2$ specific heat. (C) The temperature dependence of the $Hf_2CO_2$ thermal expansion coefficient.

**Table 1.** Parameters for calculating the thermal conductivities of $M_2CO_2$ (M=Ti, Zr, Hf) MXenes.

| MXenes | | $v_j$ $(10^3 ms^{-1})$ | | | $\gamma_j$ | | | $<\gamma_j^2>$ | | |
|---|---|---|---|---|---|---|---|---|---|---|
| | | ZA | TA | LA | ZA | TA | LA | ZA | TA | LA |
| $Ti_2CO_2$ | Armchair | 2.566 | 2.666 | 2.974 | -5.348 | 4.374 | 1.737 | 4088 | 20.42 | 5.379 |
| | Zigzag | 2.252 | 2.974 | 2.279 | -4.457 | 4.317 | 2.252 | 4089 | 19.83 | 6.705 |
| $Zr_2CO_2$ | Armchair | 2.379 | 2.518 | 2.715 | -0.573 | 2.432 | 1.231 | 19.79 | 6.155 | 2.639 |
| | Zigzag | 2.113 | 2.640 | 2.136 | -2.008 | 3.039 | 1.433 | 298.1 | 10.13 | 2.911 |
| $Hf_2CO_2$ | Armchair | 1.826 | 1.919 | 2.066 | -0.164 | -1.254 | 1.032 | 1.256 | 1382 | 2.088 |
| | Zigzag | 1.641 | 2.075 | 1.656 | -0.263 | -0.916 | 1.240 | 2.316 | 1392 | 2.391 |

**Table 2.** $Hf_2CO_2$ carrier mobility

| Carrier type | $m^*_{ex}/m_0$ | $m^*_{ey}/m_0$ | $E_{1x}$ | $E_{1y}$ | $C_x$ | $C_y$ | $\mu_x$ | $\mu_y$ |
| --- | --- | --- | --- | --- | --- | --- | --- | --- |
| | | | (eV) | | ($Jm^{-2}$) | | ($10^3\ cm^2V^{-1}s^{-1}$) | |
| e | 0.231 | 2.162 | 10.57 | 7.101 | 293.6 | 291.0 | 0.329 | 0.077 |
| h (upper) | 0.423 | 0.164 | 7.636 | 2.297 | 293.6 | 291.0 | 0.924 | 26.0 |
| h (lower) | 0.164 | 0.414 | 2.023 | 7.422 | 293.6 | 291.0 | 34.3 | 1.00 |

Carrier types "e" and "h" denote "electron" and "hole," respectively. $m^*_{ex}$ and $m^*_{ey}$ are the effective masses along the zigzag (x-) and armchair (y-) directions, respectively. $E_{1x}$ and $E_{1y}$ are the deformation potential constants, and $C_x$ and $C_y$ are the elastic moduli. $\mu_x$ and $\mu_y$ are the room-temperature carrier mobilities.

## ASSOCIATED CONTENT

**Supporting Information**. The electronic energy bands for all $M_2CO_2$ MXenes, and the corresponding thermal conductivities and carrier mobilities for $Ti_2CO_2$ and $Zr_2CO_2$, and the analyzation for the relationship between the mechanical strength and Grüneisen parameter, are listed. This material is available free of charge via the Internet at http://pubs.acs.org.

## AUTHOR INFORMATION

**Corresponding Author**

* E-mail address: dushiyu@nimte.ac.cn

**Notes**

The authors declare no competing financial interest.

## ACKNOWLEDGMENT

The authors acknowledge the support of the Division of Functional Materials and Nanodevices, Ningbo Institute of Materials Technology and Engineering, Chinese Academy of Sciences, the National Natural Science Foundation of China (Grant Nos. 51372046, 51479037 and 91226202), the Ningbo Municipal Natural Science Foundation (No.2014A610006), ITaP at Purdue University for computing resources and the key technology of nuclear energy, 2014, CAS Interdisciplinary Innovation Team.REFERENCES

(1) Yu, L., et al., *Graphene/MoS$_2$ hybrid technology for large-scale two-dimensional electronics.* Nano Letters, 2014. **14**(6): p. 3055-3063.
(2) Qiao, J., et al., *High-mobility transport anisotropy and linear dichroism in few-layer black phosphorus.* Nat Commun, 2014. **5**: p. 4475.
(3) Novoselov, K.S., *Electric field effect in atomically thin carbon films.* Science, 2004. **306**(5696): p. 666-669.
(4) Balandin, A.A., et al., *Superior thermal conductivity of single-layer graphene.* Nano Letters, 2008. **8**(3): p. 902-907.
(5) Novoselov, K.S., et al., *Two-dimensional atomic crystals.* Proceedings of the National Academy of Sciences, 2005. **102**(30): p. 10451-10453.
(6) Li, L., et al., *Black phosphorus field-effect transistors.* Nat Nano, 2014. **9**(5): p. 372-377.
(7) Fei, R., et al., *Enhanced thermoelectric efficiency via orthogonal electrical and thermal conductances in phosphorene.* Nano Letters, 2014. **14**(11): p. 6393-6399.
(8) Fei, R. and L. Yang, *Strain-Engineering the Anisotropic Electrical Conductance of Few-Layer Black Phosphorus.* Nano Letters, 2014. **14**(5): p. 2884-2889.
(9) Mak, K.F., et al., *Atomically thin MoS$_2$: A new direct-gap semiconductor.* Physical Review Letters, 2010. **105**(13): p. 136805.
(10) van der Zande, A.M., et al., *Grains and grain boundaries in highly crystalline monolayer molybdenum disulphide.* Nat Mater, 2013. **12**(6): p. 554-561.
(11) Xu, M., et al., *Graphene-like two-dimensional materials.* Chemical Reviews, 2013. **113**(5): p. 3766-3798.
(12) Partoens, B. and F.M. Peeters, *From graphene to graphite: Electronic structure around the K point.* Physical Review B, 2006. **74**(7): p. 075404.
(13) Song, L., et al., *Large scale growth and characterization of atomic hexagonal boron nitride layers.* Nano Letters, 2010. **10**(8): p. 3209-3215.
(14) Yan, R., et al., *Thermal conductivity of monolayer molybdenum disulfide obtained from temperature-dependent raman spectroscopy.* Acs Nano, 2013. **8**(1): p. 986-993.
(15) RadisavljevicB, et al., *Single-layer MoS$_2$ transistors.* Nat Nano, 2011. **6**(3): p. 147-150.
(16) Wood, J.D., et al., *Effective Passivation of Exfoliated Black Phosphorus Transistors against Ambient Degradation.* Nano Letters, 2014. **14**(12): p. 6964-6970.
(17) Naguib, M., et al., *Two-dimensional nanocrystals produced by exfoliation of Ti$_3$AlC$_2$.* Advanced Materials, 2011. **23**(37): p. 4248-4253.
(18) Naguib, M., et al., *Two-dimensional transition metal carbides.* Acs Nano, 2012. **6**(2): p. 1322-1331.23

(19) Naguib, M., et al., *New two-dimensional niobium and vanadium carbides as promising materials for Li-Ion Batteries.* Journal of the American Chemical Society, 2013. **135**(43): p. 15966-15969.
(20) Halim, J., et al., *Transparent conductive two-dimensional titanium carbide epitaxial thin films.* Chemistry of Materials, 2014. **26**(7): p. 2374-2381.
(21) Naguib, M., et al., *25th anniversary article: MXenes: a new family of two-dimensional materials.* Advanced Materials, 2014. **26**(7): p. 992-1005.
(22) Ghidiu, M., et al., *Synthesis and characterization of two-dimensional $Nb_4C_3$ (MXene).* Chemical Communications, 2014. **50**(67): p. 9517-9520.
(23) Barsoum, M.W. and M. Radovic, *Elastic and mechanical properties of the MAX phases.* Annual Review of Materials Research, 2011. **41**(1): p. 195-227.
(24) Hu, Q., et al., *Two-dimensional $Sc_2C$: A reversible and high-capacity hydrogen storage material predicted by first-principles calculations.* International Journal of Hydrogen Energy, 2014. **39**(20): p. 10606-10612.
(25) Naguib, M., et al., *MXene: a promising transition metal carbide anode for lithium-ion batteries.* Electrochemistry Communications, 2012. **16**(1): p. 61-64.
(26) Tang, Q., Z. Zhou, and P. Shen, *Are MXenes promising anode materials for Li ion batteries? Computational studies on electronic properties and Li storage capability of $Ti_3C_2$ and $Ti_3C_2X_2$ (X = F, OH) monolayer.* Journal of the American Chemical Society, 2012. **134**(40): p. 16909-16916.
(27) Lukatskaya, M.R., et al., *Cation intercalation and high volumetric capacitance of two-dimensional titanium carbide.* Science, 2013. **341**(6153): p. 1502-1505.
(28) Ghidiu, M., et al., *Conductive two-dimensional titanium carbide 'clay' with high volumetric capacitance.* Nature, 2014. **516**(7529): p. 78-81.
(29) Peng, Q., et al., *Unique lead adsorption behavior of activated hydroxyl group in two-dimensional titanium carbide.* Journal of the American Chemical Society, 2014. **136**(11): p. 4113-4116.
(30) Mashtalir, O., et al., *Intercalation and delamination of layered carbides and carbonitrides.* Nat Commun, 2013. **4**: p. 1716.
(31) Gan, L.-Y., et al., *First-principles analysis of $MoS_2/Ti_2C$ and $MoS_2/Ti_2CY_2$ (Y=F and OH) all-2D semiconductor/metal contacts.* Physical Review B, 2013. **87**(24): p. 245307.
(32) Ma, Z., et al., *Tunable band structures of heterostructured bilayers with transition-metal dichalcogenide and MXene monolayer.* The Journal of Physical Chemistry C, 2014. **118**(10): p. 5593-5599.
(33) Xie, Y., et al., *Role of surface structure on Li-ion energy storage capacity of two-dimensional transition-metal carbides.* Journal of the American Chemical Society, 2014. **136**(17): p. 6385-6394.
(34) Eames, C. and M.S. Islam, *Ion Intercalation into two-dimensional transition-metal carbides: global screening for new high-capacity battery materials.* Journal of the American Chemical Society, 2014. **136**(46): p. 16270-16276.
(35) Gan, L.-Y., D. Huang, and U. Schwingenschlögl, *Oxygen adsorption and dissociation during the oxidation of monolayer $Ti_2C$.* Journal of Materials Chemistry A, 2013. **1**(43): p. 13672.
(36) Zha, X.-H., et al., *Role of the surface effect on the structural, electronic and mechanical properties of the carbide MXenes.* EPL (Europhysics Letters), 2015. **111**: p. 26007.

(37) Kresse, G. and J. Furthmüller, *Efficient iterative schemes for ab initio total-energy calculations using a plane-wave basis set.* Physical Review B, 1996. **54**(16): p. 11169-11186.
(38) Perdew, J.P., K. Burke, and M. Ernzerhof, *Generalized gradient approximation made simple.* Physical Review Letters, 1996. **77**(18): p. 3865-3868.
(39) Heyd, J., G.E. Scuseria, and M. Ernzerhof, *Hybrid functionals based on a screened Coulomb potential.* The Journal of Chemical Physics, 2003. **118**(18): p. 8207-8215.
(40) Paier, J., et al., *Screened hybrid density functionals applied to solids.* The Journal of Chemical Physics, 2006. **124**(15): p. 154709.
(41) Blöchl, P.E., *Projector augmented-wave method.* Physical Review B, 1994. **50**(24): p. 17953-17979.
(42) Turney, J.E., et al., *Predicting phonon properties and thermal conductivity from anharmonic lattice dynamics calculations and molecular dynamics simulations.* Physical Review B, 2009. **79**(6).
(43) Momma, K. and F. Izumi, *VESTA 3 for three-dimensional visualization of crystal, volumetric and morphology data.* Journal of Applied Crystallography, 2011. **44**(6): p. 1272-1276.
(44) Togo, A., F. Oba, and I. Tanaka, *First-principles calculations of the ferroelastic transition between rutile-type and CaCl$_2$-type SiO$_2$ at high pressures.* Physical Review B, 2008. **78**(13): p. 134106.
(45) Gonze, X. and C. Lee, *Dynamical matrices, Born effective charges, dielectric permittivity tensors, and interatomic force constants from density-functional perturbation theory.* Physical Review B, 1997. **55**(16): p. 10355-10368.
(46) Balandin, A.A., *Thermal properties of graphene and nanostructured carbon materials.* Nature Materials, 2011. **10**(8): p. 569-581.
(47) Klemens, P.G. and D.F. Pedraza, *Thermal conductivity of graphite in the basal plane.* Carbon, 1994. **32**(4): p. 735-741.
(48) Klemens, P.G., *Theory of Thermal Conduction in Thin Ceramic Films.* International Journal of Thermophysics, 2001. **22**(1): p. 265-275.
(49) Mounet, N. and N. Marzari, *First-principles determination of the structural, vibrational and thermodynamic properties of diamond, graphite, and derivatives.* Physical Review B, 2005. **71**(20): p. 205214.
(50) Andrew, R.C., et al., *Mechanical properties of graphene and boronitrene.* Physical Review B, 2012. **85**(12): p. 125428.
(51) Grimme, S., *Semiempirical GGA-type density functional constructed with a long-range dispersion correction.* Journal of Computational Chemistry, 2006. **27**(15): p. 1787-1799.
(52) Pu, H.H., et al., *Anisotropic thermal conductivity of semiconducting graphene monoxide.* Applied Physics Letters, 2013. **102**(22): p. 223101.
(53) Bottger, P.H.M., et al., *Influence of Ball-Milling, Nanostructuring, and Ag Inclusions on Thermoelectric Properties of ZnSb.* Journal of Electronic Materials, 2010. **39**(9): p. 1583-1588.
(54) Upadhyay Kahaly, M. and U.V. Waghmare, *Size dependence of thermal properties of armchair carbon nanotubes: A first-principles study.* Applied Physics Letters, 2007. **91**(2): p. 023112.
(55) Zha, X.-H., R.-Q. Zhang, and Z. Lin, *Point defect weakened thermal contraction in monolayer graphene.* The Journal of Chemical Physics, 2014. **141**(6): p. 064705.

(56) Bruzzone, S. and G. Fiori, *Ab-initio simulations of deformation potentials and electron mobility in chemically modified graphene and two-dimensional hexagonal boron-nitride.* Applied Physics Letters, 2011. **99**(22): p. 222108.

(57) Khazaei, M., et al., *Novel electronic and magnetic properties of two-dimensional transition metal carbides and nitrides.* Advanced Functional Materials, 2013. **23**(17): p. 2185-2192.

(58) Heyd, J., et al., *Energy band gaps and lattice parameters evaluated with the Heyd-Scuseria-Ernzerhof screened hybrid functional.* Journal of Chemical Physics, 2005. **123**(17): p. 174101.

(59) Lide D R, e., *CRC Handbook of chemistry and physics (84th ed.) CRC Press. Boca Raton, Florida.* 2003.

(60) Khazaei, M., et al., *Two-dimensional molybdenum carbides: potential thermoelectric materials of the MXene family.* Physical Chemistry Chemical Physics, 2014. **16**(17): p. 7841.

(61) Eklund, P., et al., *The $M_{n+1}AX_n$ phases: Materials science and thin-film processing.* Thin Solid Films, 2010. **518**(8): p. 1851-1878.

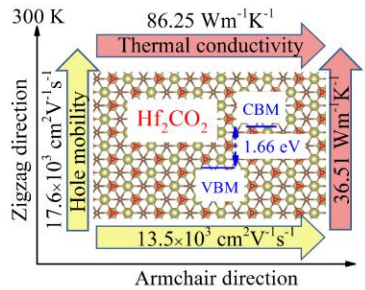

Table of Contents Graphic